# Irreversible Dynamics of Abrikosov Vortices in Type-two Superconductors

S. Pace[1,2], G. Filatrella[1,3], G.Grimaldi[1,2], A. Nigro[1,2]

[1]INFM, Regional Laboratory of Salerno,Via S. Allende I-84081 Baronissi, Italy

[2]Dept. of Physics "E.R. Caianiello", University of Salerno, Via S. Allende I-84081 Baronissi, Italy

[3]Dept. of Biological and Environmental Sciences, Via Port'Arsa 11, I-82100 Benevento, Italy

**Abstract**

The voltage shift $R_f I_c$ with respect to the flux flow ohmic behavior of the current-voltage characteristic in type II superconductors is ascribed to the irreversible processes occurring when a vortex crosses defects. We include irreversible effects of the vortices-defects interaction into an effective law of motion. The obtained current-voltage characteristic at finite temperature is in agreement with experimental data.

Corresponding author:
Sandro Pace
INFM Regional Laboratory of Salerno
Physics Department
v. S. Allende
I-84081 Baronissi (SA)
Italy
Phone: +39-089965369
Fax: +39-089965275
e-mail: pace@sa.infn.it

## 1. Introduction

In the mixed state of type II superconductors the mechanism underlying the vortex-defect interaction plays a fundamental role in determining the transport properties as well as the magnetic behavior of actual superconducting samples [1-3]. Magnetic field and temperature dependences of resistance and critical current have been deeply studied, but important features of current transport properties have not yet been completely explained. For instance the appearance of quasi-static magnetization cycles implies the existence of intrinsic irreversible processes induced by the flux quanta motion, whose general microscopic origin has not yet been clearly identified. Another common and unexplained feature observed in low temperature superconductors, is the linear dependence of the electric field $E \sim (J-J_c)$, above the critical current density, $J_c$ [4-6]. Also in high temperature superconductors a shifted linear behavior in the current-voltage (I-V) characteristics seems to be asymptotically reached for currents much higher than $I_c$. As an example, Fig.1 shows the experimental I-V curves at 4.2K of a low critical temperature ($T_c$) homogeneous material and of a superconducting tape, and of high temperature superconducting (HTS) films at 77K. Above the critical current, $I_c$, a linear dependence $V=R_f (I-I_c)$ is observed with the presence of a voltage shift $\Delta V= R_f I_c$, where $R_f$ is the measured flux flow resistance [7,8]. Due to inhomogeneities the I-V characteristics of HTS samples do not show in a clear way the linear dependence, and for $J \cong J_c$ they can be approximated by a power law dependence $V \approx I^n$, where the index $n$ characterizes dissipations: the higher is $n$ the lower are thermal induced losses below $J_c$ [6,9,10]. A similar behavior appears in other HTS samples [11].

Macroscopic dissipations occur in type II superconductors because Abrikosov vortices move under the effect of a current density $\mathbf{J}$ flowing through the sample; the current drives each flux quantum $\mathbf{\Phi_0}$ via a Lorentz-like force per unit length $\mathbf{f_L}=\mathbf{J}\mathsf{X}\mathbf{\Phi_0}$. Consequently for each moving vortex, a friction force $-\eta \cdot \mathbf{v}$ proportional to the vortex velocity appears, due to the normal current produced by the electric field arising from the motion of the vortex structure [1]. The balance between the Lorentz force and the friction force determines the equilibrium velocity, which is then proportional to the voltage. As a straightforward consequence the I-V characteristic would be expected linear: $\mathbf{E}= \rho_f\mathbf{J}$, where $\rho_f$ is the flux flow resistivity. Deviations of the I-V characteristic from the ohmic behavior $\mathbf{E}=\rho_f\mathbf{J}$ are ascribed to the presence of material defects, such as impurities or spatial inhomogeneities of the superconducting parameters. Indeed defects make the free energy $F$ of a single flux quantum a function of the vortex position, so that metastable equilibrium configurations correspond to $F$ minima. The height $\Delta F$ of the well is determined by the energy necessary to create the vortex normal core when the flux quantum is driven outside the defect. Defects are then treated as potential wells, whose gradient determines the pinning force $\mathbf{f_p}$. For a weak external driving current, vortices cannot move and are pinned by the force $f_p$. Only for $f_L>f_p$, i.e. $J>J_c$, is the vortex driven outside the defect in a flux flow state. Above $J_c$, the flux flow linear branch intersects $J_c$, as shown in Fig. 1; a proposed explanation for this behavior is the existence of a dynamic pinning force equal to the static one and always opposite to the vortex motion [12,13]. The latter assumption is in contrast with the standard potential well description of the pinning center. In this Letter we propose that the contradiction can be overcome by taking into account the irreversible non equilibrium processes that occur in the vortex motion through defects. We will also discuss the consequences of the modified single vortex dynamics on the collective vortices motion [14].

The paper is organized as follows: in Section 2 we outline the physical mechanism leading to the effective law of motion for the vortices; in Section 3 we derive the I-V characteristic from the Langevin equation. As usual discussions and conclusions are drawn in the last Section.

## 2. The Langevin equation of the effective vortex motion

The essential point is that rather than being always opposite to the motion, any force due to a potential well should accelerate vortices falling down into the well, and decelerate vortices going up. For this reason, if the external current is sufficiently higher than $J_c$, the vortex average velocity should be unaffected by the presence of a potential well and it should be equal to the flux flow velocity, leading to a linear I-V dependence intersecting the origin. Only for currents just above the critical value should the potential well cause substantial changes in this velocity. As an example an analytic description of the motion in a potential well can be performed, assuming the well-known washboard potential with a damping force proportional to the velocity [15,16]; for $J \gg J_c$, a linear I-V characteristic intersecting the origin is obtained. This result is independent of the particular shape of the potential well [17], so that the presence of the experimental voltage shift cannot be explained by analyzing the motion of independent vortices through potential wells. To explain the voltage shift and to overcome the contradiction of the usual description we suggest to consider irreversible non equilibrium processes that occur when a vortex crosses defects. We recall that it has already been suggested that non-equilibrium phenomena play a role in the flux flow motion even in the absence of defects [18,19]. Indeed, in non-equilibrium conditions, the work done by a system is always less than its free energy variation so that the free energy cannot be considered like a potential energy. The vortex falling into the defect is not a quasi-static process, and consequently the free energy difference of the vortex inside and outside the defect, cannot be treated as a potential well. While entering in the defect the actual heat loss due to excess entropy is larger than the corresponding reversible process heat loss, which in the case of a non superconducting defect is determined by the entropy reduction generated by the disappearing of the vortex normal core. The microscopic origin of this additive dissipation is the condensation energy of quasi particles into Cooper pairs, which produces thermal phonon emission [20]. Moreover, in the reversible exit of the vortex from the defect, a work equal to the free energy variation should be done and, in the same time, the vortex would then absorb the heat due to the entropy variation determined by the normal core formation. In dynamic non-equilibrium conditions the work for the vortex exit could be even larger than the free energy variation.

We therefore propose an effective equation for a one-dimensional single vortex motion to take into account the irreversible process of a vortex entry into a defect. The starting point is nevertheless the free energy change $\Delta F$ of the vortex when trapped by a pinning center. The shape of the corresponding potential is not known in detail, but to a first order approximation, we just assume a triangular profile as shown in Fig.2a. The bias current introduces a Lorentz force in the $x$ direction and adds a linearly decreasing potential $U'(x) = -f_L \cdot x$, as shown in Fig. 2b. In order to account for the irreversible processes discussed above, an effective potential can be retained as shown in Fig. 2c. In fact, starting from the bottom of a pinning center ($x_0$) a potential barrier must be overcome to allow the vortex to go out. Nevertheless, when the vortex is outside the well ($x_0 + b < x < x_0 + L$), the Lorentz force only drives the vortex towards the next center of a defect. Indeed, since the entry into the nearest defect is an irreversible process, there is no additional force related to the free energy change, which is dissipated as heat. This phenomenon is analogous to the expansion of a gas closed in a cylinder by a piston [18]: If a sudden expansion is obtained by an external force acting on the piston, the gas will not help the piston motion, and no force exerted by the gas expansion has to be considered. Similarly we assume negligible the force acting on the vortex while entering into the defect.

The process described in Fig. 2c can account for an I-V curve shifted respect to the origin rather than one asymptotically reaching V=R$_f$I as results from a purely periodic potential. We note that if the irreversible effects would be included as an additional friction, as was earlier suggested in Ref [18], this would only be reflected in a different slope of the I-V curve; rather, one has to treat the details of the interaction between the vortex and the pinning center. To such purpose we describe the process displayed in Fig. 2c with the following dynamical equation:

$$\eta \frac{dx}{dt} = F_L + \xi(t) \qquad\qquad x_0 + b < x \le x_0 + L,$$

$$\eta \frac{dx}{dt} = (F_L - F_p) + \xi(t) \qquad x_0 \le x \le x_0 + b.$$

$$(1)$$

Here $F_L$ and $F_p$ are the forces exerted on a single vortex by the bias current and the pinning center, respectively, $\eta$ is the damping, $\xi$ the stochastic term due to thermal fluctuations whose correlator is $<\xi\xi'> = 2\eta K_B T \delta(t-t')$, $L$ is the distance between two pinning centers and $b$ is the effective size of the defect. When the vortex reaches the next pinning center, its equation of motion is again approximated by a Langevin equation of the type (1). In other words, in the effective Langevin equation the pinning force f$_p$ is not periodic – it must be included only when the motion is against the Lorentz force.

## 3. I-V characteristics at finite temperature

Since the main problem is to find out the time a vortex spends to reach the next pinning center, a natural framework to tackle the problem is to compute the Mean First Passage Time (MFPT) $t$ for the basic process of a vortex starting from the bottom of a pinning center to reach the next one. More precisely, we assume that there is a reflecting barrier in $x_0$ and an absorbing barrier in $x_0 + L$ to compute the average time to move forward. The formal solution for the MFPT is thus obtained solving the double integral [16,21]:

$$t = \frac{\eta}{k_B T} \int_{x_0}^{x_0+L} dy \exp\left[\frac{U(y)}{k_B T}\right] \int_{x_0}^{y} dz \ \exp\left[\frac{-U(z)}{k_B T}\right],$$

$$(2)$$

where $U$ is the potential that gives rise to the forces of Eq. (1). We see here the advantage of having assumed that the pinning force is piecewise constant: the double integral (2) can be analytically evaluated and the solution reads:

$$
\begin{aligned}
t_{forward} &= \frac{\eta b}{\left(F_p - F_L\right)}\left[\frac{k_b T}{\left(F_p - F_L\right)b}\left(e^{\left(F_p - F_L\right)b/k_B T} - 1\right) - 1\right] \\
&+ \frac{\eta(L-b)}{F_L}\left[\frac{k_b T}{F_L(L-b)}\left(e^{-F_L(L-b)/k_b T} - 1\right) - 1\right] + \\
&\frac{\eta k_b T}{F_L\left(F_L - F_p\right)}\left[\left(1 - e^{\left(F_p - F_L\right)b/k_B T}\right)\left(1 - e^{-F_L(L-b)/k_B T}\right)\right]
\end{aligned}
\tag{3}
$$

There is also a finite probability for the vortex to jump backwards, i.e. against the Lorentz force. The corresponding MFPT $t_{backwards}$ can be analogously computed imposing an absorbing barrier in $x_0 - L$ for the backward process, and so one finally retrieves the average velocity:

$$
<v> = L \cdot \left(1/t_{forward} - 1/t_{backward}\right)
\tag{4}
$$

Eq.s (3,4) are the basic result of this Letter: they state that the vortex average velocity resulting from Eq.(1) can be estimated also at finite temperature. It is convenient to normalize distances respect to the distance $L$ between two defects, time respect to $\eta/(k_B T)$, forces respect to the pinning force $F_p$, and to introduce the normalized temperature $\theta = k_B T / F_p L$. We have reproduced the current ($I \propto F_L$) – voltage ($V \propto <v>$) characteristic in Fig. 3. From the figure it is clear that the I-V characteristic can be approximated for $I \gg I_c$ by V=R$_f$(I-Ic). One can find the asymptotic solution for the noiseless system (1) [$\xi(t) = 0$] observing that the velocity, to the leading order in $(b/L) \cdot (F_p/F_L)$ is $<v> \cong 1/\eta\left(F_L - bF_p/L\right)$, or, in terms of the I-V characteristic V$\cong$R$_f$(I-$b$ I$_c$/ $L$). This accounts for the voltage shift of the linear branch in Fig. 3. On the contrary, the usual washboard potential gives the asymptotic behavior V=R$_f$I as shown in Fig.3 (squares).

In Fig. 4 we show a detailed temperature dependence of the I-V characteristics. We note that the curves clearly exhibit a voltage shift at all temperatures. Moreover, the critical current density value might be underestimated by the linear extrapolation of the experimental data [6].

## 4. Discussion and conclusion

The above analysis regards the basic single vortex interaction with defects and the consequent dynamic processes. Actually the vortex physics is much more complicated due to many interactions such as: a) a single vortex with many pinning centers, b) vortices among them, c) a vortex lattice with pinning centers. The collective pinning theories include these effects and determine effective pinning potentials for flux bundles [14], producing a complex behavior such as the irreversibility line H$_{irr}$(T), above which hysteretic behavior disappears [22-26]. However the I-V characteristics derived within the collective pinning approach do not account for the observed shift above $I_c$ [27-32]. In a collective pinning approach vortex dynamics has been computed [30,33,34]. The overall effect of the interaction among vortices and pinning sites amounts to an effective force on each vortex opposite to the motion - analogous to the force due to the irreversible processes here considered. However, the aforementioned collective pinning approaches employ a microscopic perturbation technique that, starting from an unperturbed flux flow velocity v, results in a relative change of the vortex velocity $\delta v / v$ that approaches zero for current densities much larger than $J_c$ [30,33,34]. (The same approaches explain microscopically the existence of a net pinning force by means of a divergence of the ratio $\delta v / v$ for J below $J_c$.) The main point is that, as in the single vortex dynamics, also in the collective motion case by increasing the external current pinning forces due to potential wells average out at zero value. For $J \gg J_c$ the vortex array becomes a rigid lattice [31], thus time average and spatial average of the pinning forces should result equal to zero [32]. As a consequence, in the presence of potential wells, also in the collective pinning theory the I-V characteristics should be straight lines crossing the origin with the $R_f$ slope, which is not the case for $J \gg J_c$ (see Fig. 1). On the contrary in our approach the I-V curve shifted respect to the origin results as the asymptotic high current solution of a single vortex dynamics or, equivalently, of a chain of many diluted vortices [17].

The forces included in Eq. (1) will also affect the collective vortices motion. Assuming that the interacting vortices move equally spaced, as in the widely employed approach of Koshelev and Vinokur in the limit of high velocity (i.e., $F_L \gg F_p$), the total force acting on the chain is the average force of the pinning centers [29,31,32]. The forces in Eq. (1) *do not* average to 0 as they would in any periodic potential: We therefore anticipate that our approach will give a shifted linear behavior of the I-V characteristic for the collective motion *at all velocities*. In the low drive limit ($I \equiv I_c$), where the vortex-vortex interaction cannot be assumed to be rigid, the effects of spatial fluctuations of the inhomogeities are expected to add to the thermal fluctuations, possibly giving rise to a melting of the vortex chain. How the melting of vortices arises in the presence of irreversible processes such as those considered in this work is at the moment an open question.

In summary, we have modified the usual approach to treat the pinning force arising from the presence of defects in order to include irreversible processes into an effective equation of motion. This leads to an analytical expression for the I-V curve also at finite temperature. We have showed that the resulting I-V is shifted with respect to the ohmic straight line, in agreement with the experiments. Finally we remark that the microscopic irreversible mechanism of vortex-defect interaction could give a deeper understanding of vortex motion and of the derived energy losses. We hope that such understanding can provide a way forward for improving the transport properties in type II superconductors.

We thank A. M. Testa, P. Caputo, and C. Senatore for helpful suggestions.

**Figure Captions**

Fig. 1. Current-voltage characteristics of type-II superconducting samples . Full squares have been measured on a YBCO thin film (T=77K, in self field), open circles have been recorded on a YBCO thick film deposited on a metallic tape (T=77K, in self field). Triangles data are from Ref. [8] for a Nb foil (T=4.22K, with an externally applied field of 0.1 Tesla), crosses are from Ref. [10] for a NbTi strand (T=4.2K, with an externally applied field of 5 Tesla). Current data are normalized to the critical current values.

Fig. 2. Schematic single vortex (full circle) motion through defects. **a)** The triangular profile of the standard potential in the absence of bias current. **b)** The tilted potential due to the bias current. **c)** The effective potential used in our model in order to account for the irreversible processes discussed in the text. The dotted line corresponds to the vortex entry into the nearest defect (irreversible process) where the vortex is driven only by the Lorentz force.

Fig. 3. Current-voltage characteristic obtained from Eq.s (3-4) (circles) and the standard washboard potential (squares) compared with the ohmic straight line (dashed line). The solid line shows the asymptotic noiseless solution. Parameters of the simulations: distance between two defects $L = 2b$, normalized temperature $\theta$=0.02.

Fig. 4. Detail of the current-voltage characteristic obtained from Eq.s (3,4) at several temperatures. Solid lines refer to different normalized (respect to $F_p L/k_B$) temperatures $\theta = 0.10, 0.08, 0.06, 0.04, 0.02$ (clockwise). The dashed line denotes the ohmic straight line, and the dotted line is the analytic noiseless solution for $\theta = 0$. The distance between two defects is $L = 2b$.

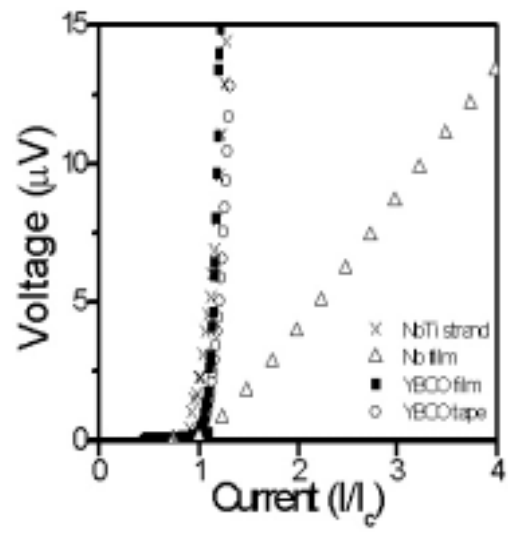

**Fig. 1** S. Pace et al.

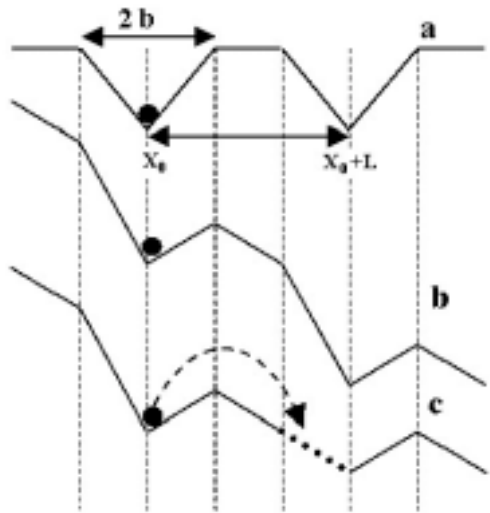

**Fig. 2** S. Pace et al.

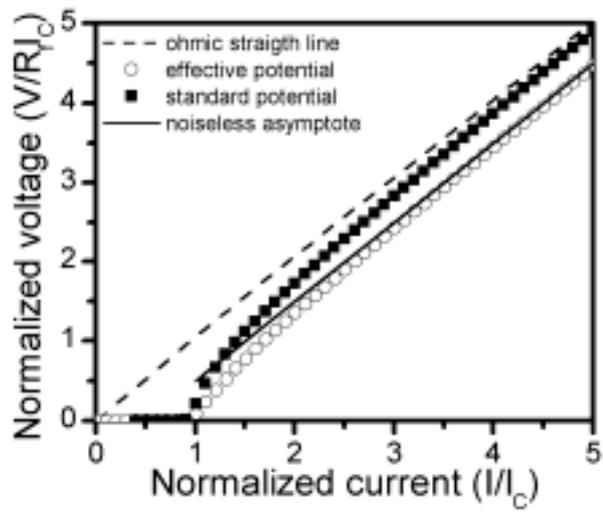

**Fig. 3** S. Pace et al.

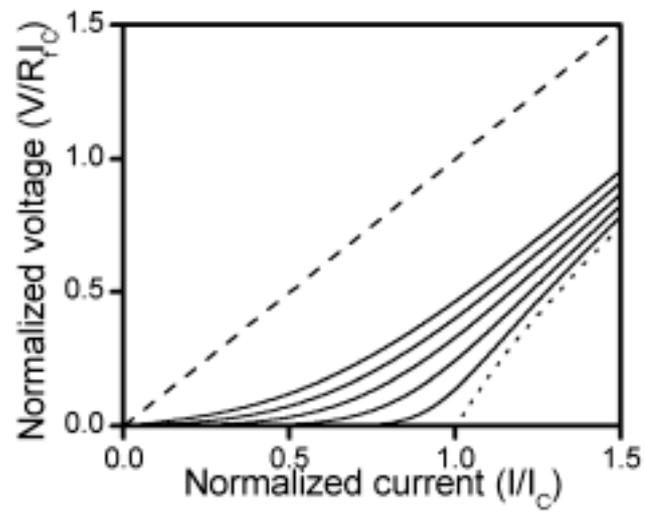

**Fig. 4** S. Pace et al.